\documentclass{article}
\usepackage{graphicx}
\usepackage[margin=1.0in]{geometry}
\usepackage{comment}
\usepackage{multirow}
\usepackage{hyperref}
\usepackage{indentfirst}

\usepackage{graphicx, amsmath, amsthm, layout, epstopdf}
\usepackage{textcomp}
\usepackage{ amssymb }
\usepackage[labelfont=bf]{caption}
\usepackage{subcaption}
\usepackage{etoolbox}
\usepackage{listings}
\usepackage[framed,numbered,autolinebreaks,useliterate]{mcode}
\usepackage{url,textcomp}
\usepackage{multicol}
\usepackage{multirow}
\usepackage{algorithmic}
\usepackage[linesnumbered]{algorithm2e}

\usepackage{color}

\makeatletter
\patchcmd{\chapter}{\if@openright\cleardoublepage\else\clearpage\fi}{}{}{}
\makeatother

\usepackage[ansinew]{inputenc}

\usepackage{ifpdf}		
\usepackage{float}
\usepackage{amsmath}
\usepackage{enumerate}
\usepackage{listing_style}

\begin{document}

\title{Parameter inference method for stochastic single-cell dynamics from tree-structured data}
\author{Irena Kuzmanovska, Andreas Milias-Argeitis, Christoph Zechner and Mustafa Khammash}

\maketitle


\begin{abstract}
With the advance of experimental techniques such as time-lapse fluorescence microscopy, the availability of single-cell trajectory data has vastly increased, and so has the demand for computational methods suitable for parameter inference with this type of data. However, most of currently available methods treat single-cell trajectories independently, ignoring the mother-daughter relationships and the information provided by population structure. This information is however essential if a process of interest happens at cell division, or if it evolves slowly compared to the duration of the cell cycle. In this work, we highlight the importance of tracking cell lineage trees and propose a Bayesian framework for parameter inference on tree-structured data. Our method relies on a combination of Sequential Monte Carlo for likelihood approximation and Markov Chain Monte Carlo for parameter sampling. We demonstrate the capabilities of our inference framework on two simple examples in which the lineage tree information is necessary: one in which the cell phenotype can only switch at cell division and another where the cell type fluctuates randomly over timescales that extend well beyond the lifetime of a single cell.
\end{abstract}


\section{Introduction}
\label{sec:introduction}

Biochemical processes in isogenic cells exhibit substantial heterogeneity \cite{Mcadams1997, Elowitz2002}. Understanding the latter demands for experimental techniques that can resolve such processes at the single-cell level. In contrast to bulk measurements, these techniques provide not only access to the average behavior of intracellular dynamics, but also its variability across cells and over time. Most single-cell techniques, however, reveal only very few components simultaneously that are often multiple steps away from the actual quantities of interest. The dynamics of a promoter, for instance, may not be accessible directly, but only indirectly through a fluorescent reporter that is expressed upon activation of this promoter \cite{Hansen2013}. Statistical inference in combination with mathematical models provides a means to reconstruct inaccessible parameters from available measurements, making it instrumental for studying biochemical processes based on single-cell data.

How such inference can be performed depends strongly on the way the data has been collected: flow cytometry measurements, for instance, reveal fluorescence values across a population but individual cells cannot be tracked over time. Consequently, measurements at two different time instances are considered statistically independent. Time-lapse microscopy techniques permit tracking of single cell trajectories over the duration of a whole experiment \cite{time-lapse}, which in turn provides a handle also on the temporal correlation of the underlying process. This additional degree of information can dramatically improve the inference of unknown process parameters \cite{Zechner2014}.

Most existing inference approaches consider single cell trajectories to be statistically independent of each other \cite{Batt2015, Zechner2014, Hansen2013, wilkinson2011}. This way, however, important information stemming from the ancestry of a cell are lost: shortly after cell division, for example, two daughter cells are likely to exhibit substantial correlations, which cannot be captured by a model that assumes independence among cells. This can yield incomplete and biased results, especially when the time scale of the process under study is slow compared with the cell cycle duration.


In addition, stochastic processes of interest such as epigenetically regulated phase variation in bacteria are often driven by DNA replication just before cell division. Examples in this category are the regulation of \emph{agn43}  \cite{marjan,lim} \cite{lim} and  \emph{Pap} \cite{pap-pili,o1985gal} systems in \emph{E.coli}, and the glucosyltransferase (\emph{gtr}) gene cluster in \emph{Salmonella} \cite{salmonella}. Due to the non-reversibility of the epigenetic modifications, gene replication (and consequently cell division) is crucial for switching to happen. Therefore, cell lineage information has to be taken into account in single-cell studies of these systems.

At present, there exist only little work on the statistical inference of tree-based single-cell data. In \cite{hasenauer}, the authors have proposed a method for inferring process parameters based on approximate Bayesian computation (ABC) from single-cell trajectories. While their approach applies in principle also tree-structured data, it requires all trajectories to have the same length and resolution. The method is therefore limited to cases where cell division times exhibit negligible variability across cells and over time.


In this work, we propose a Bayesian parameter inference framework for inferring parameters from general tree-structured data. Our  method works also when the mother and daughter cells have different lifespan and different measurement points throughout their the window of acquisition. In Section \ref{sec:problem_description} we will give a mathematical description of the inference problem and the class of models we consider, in Section \ref{methods} we will present our method in details and in Section \ref{sec:applications} we will demonstrate our method on two different model examples by inferring some of their parameters.\\

\section{Problem description}
\label{sec:problem_description}

\subsection{The model}\label{model}

\begin{figure}[H]
    \centering
    \includegraphics[width=5.0in]{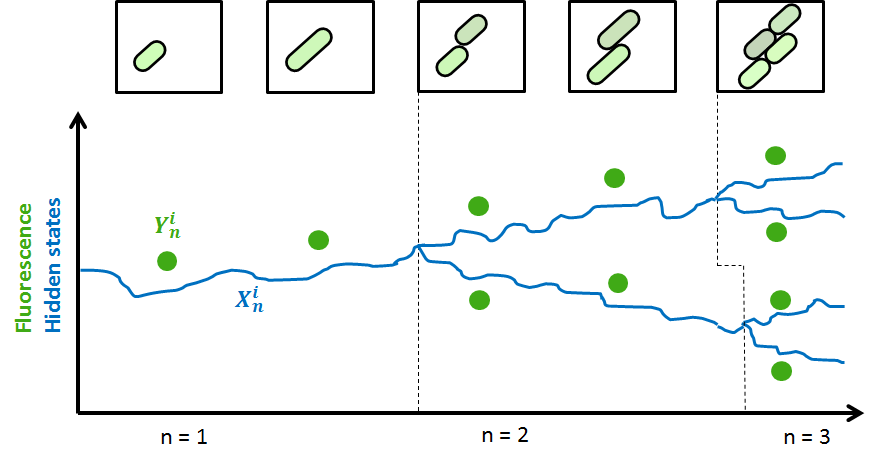}
    \caption{Graphical illustration of time-lapse microscopy images of a growing \emph{E.coli} colony which contains a fluorescent reporter gene. After the cells are segmented and tracked, the fluorescence intensity in each of them can be extracted, giving rise to a fluorescence tree-structured dataset (greed dots). The continuous blue curve represents the unobserved state trajectory of each individual cell.}
    \label{fig:drawing}
\end{figure}

To introduce the inference problem and the class of models considered here, we refer to the illustration in Figure \ref{fig:drawing}. Let us consider an intracellular biochemical process of interest modeled by a continuous-time dynamical $S$. The system behavior within each cell can be monitored with the help of a dynamic readout, such as the abundance of a fluorescent reporter protein. Through time-lapse microscopy, we assume that a growing population of single-cells and their progeny can be tracked over time and measured at multiple time points (green dots in Figure \ref{fig:drawing}), giving rise to a hierarchical tree data structure that describes the time evolution of the population.

We assume that each tree starts with a single mother at generation $0$ and that the population is followed over $N$ generations, giving rise to a total number of $2^N$ cells at the final generation.  The system $S$ describes the evolution of a set of internal states $x$ (blue curve in Figure \ref{fig:drawing}). These states can be accessed partially at discrete time points through experimental techniques yielding a corresponding readout $y$. Each cell is assigned a separate time index and a separate time of division, $T$, which can either be assumed known from the single-cell tracking data, or be inferred based on this data. We denote by $\boldsymbol{X}$ the the whole trajectory $\{x(t),t\in[0,T]\}$ from the time of birth of a cell (at $t=0$) until its division (at $t=T$). The dynamics of $S$ may evolve on a continuous, discrete or hybrid space, and similarly be stochastic, deterministic, or involve components of both types. In any case, we assume that $S$ depends on a set of constant parameters $\Theta$, which are assumed to be the same across the population.

At generation $n$ there are $2^n$ cells in the population. From this point on, we will distinguish each cell by its generation number, $n$, and an index $i$, that ranges from 1 to $2^n$. The $i^{th}$ cell of the $n^{th}$ generation gives rise to two daughters, indexed by $2i-1$ and $2i$, in generation $n+1$. Henceforth, all quantities related to a certain cell in a given lineage will be indexed by these two numbers.

Following this notation, we denote by $\boldsymbol{X}_n^i$ the state trajectory of the $i^{th}$ cell in the $n^{th}$ generation; that is, \[\boldsymbol{X}_n^i:= \{x_n^i(t),t\in[0,T_n^i]\}.\] The state trajectories of the daughters originating from a mother cell with state trajectory $\boldsymbol{X}_n^i$ will therefore be denoted by $\boldsymbol{X}_{n+1}^{2i-1}$ and $\boldsymbol{X}_{n+1}^{2i}$ respectively. The corresponding discrete set of measurements associated with $\boldsymbol{X}_n^i$ is denoted by $\boldsymbol{Y}_n^i$. More specifically, \[\boldsymbol{Y}_n^i:=\{y_n^i(t_{n,k}^i),k=1,\dots,K_n^i)\}.\]
This notation reflects the fact that the $i^{th}$  cell of the $n^{th}$ generation is observed at a total number of $K_n^i$ time points (each denoted by $t_{n,k}^i$) during its lifetime, and that the number and location of observation time points will in general be different for every cell.

We will further denote by $ P(x_{n+1}^{2i-1}(0),  x_{n+1}^{2i}(0) | x_n^i(T_n^i), \Theta )$ the distribution of the daughter initial conditions given the state of the mother just before division, and call this the \emph{transition probability} from one generation to the next. It is reasonable to assume that, once their respective initial conditions are determined based on their mother cell, the two daughters evolve independently of each other. As defined above, the transition probability mechanism may itself contain unknown parameters that need to be estimated from the data.

\subsection{The inference problem}\label{inference}
Our goal is to infer the posterior distribution of $\Theta$ given (1) the set of measured cellular readouts over the whole lineage, (2) our prior knowledge about $\Theta$ encoded in a prior distribution $\pi(\Theta)$ and (3) a measurement noise model that describes the likelihood of observing $y_n^i(t)$ given $x_n^i(t)$ (possibly also depending explicitly on unknown parameters contained in $\Theta$). The latter is given by the density $f(y_n^i(t)|x_n^i(t),\Theta)$. With this measurement model, and assuming that measurements at individual time points are independent from each other, the likelihood of the whole measurement set for a single cell can be defined as
\[P( \boldsymbol{Y}_n^i \mid \boldsymbol{X}_n^i,\Theta )=\prod_{k=1}^{K_n^i} f(y_n^i(t_{n,k}^i)|x_n^i(t_{n,k}^i),\Theta).\]

Setting \[\boldsymbol{X}_{tree}:=\{\boldsymbol{X}_n^i,i=1,\dots,2^n,~n=0,\dots,N\}\] and \[\boldsymbol{Y}_{tree}:=\{\boldsymbol{Y}_n^i,i=1,\dots,2^n,~n=0,\dots,N\},\] the joint distribution over states and measurements over a tree starting from a single individual can be written as

\begin{equation}\label{joint_likelihood}
\begin{split}
P(\boldsymbol{X}_{tree},\boldsymbol{Y}_{tree} \mid \Theta)&=P(\boldsymbol{X}_0^1 \mid \pi_x)P(\boldsymbol{Y}_0^1 \mid \boldsymbol{X}_0^1,\Theta) \\
&\times \prod_{n=1}^N \Bigg[ \left( \prod_{i=1}^{2^n-1} P(\boldsymbol{X}_n^{2i-1},\boldsymbol{X}_n^{2i} \mid \boldsymbol{X}_{n-1}^i,\Theta) \right) \left(\prod_{i=1 }^{2^n-1} P( \boldsymbol{Y}_n^i \mid \boldsymbol{X}_n^i,\Theta) \right) \Bigg],
\end{split}
\end{equation}
where $\pi_x$ is the initial distribution of $x_0^1(0)$. The likelihood of the measured outputs given $\Theta$ can therefore by obtained by marginalization of \eqref{joint_likelihood} over all possible unobserved states:
\begin{equation}\label{output_likelihood}
P(\boldsymbol{Y}_{tree} \mid \Theta)=\int P(\boldsymbol{X}_{tree},\boldsymbol{Y}_{tree} \mid \Theta)d\boldsymbol{X}_{tree}.
\end{equation}

As can be seen from the above equations, an additional difficulty of our inference problem in comparison to inference based on independent cell trajectories, is the fact that the likelihood $P(\boldsymbol{X}_{tree},\boldsymbol{Y}_{tree} \mid \Theta)$ does not factorize over the readouts of individual cells, since the tree structure of the population introduces dependencies among the  observations coming from different generations. The dependencies are generated through the unobserved state dynamics, which must therefore be taken into account.

Moreover, due to the dependencies introduced by the tree structure of the population, the integral in \eqref{output_likelihood} is analytically intractable already for very simple state dynamics and its numerically evaluation scales exponentially with the number of generations in the tree. To address these difficulties, we employ a sequential Monte Carlo (SMC) scheme as described below to approximate the marginal likelihood \eqref{output_likelihood}.

\section{Methods}
\label{methods}

\subsection{Recursive likelihood and state posterior propagation}
\label{subsec:RecLikelihood}
The joint distribution over states and observations given by \eqref{joint_likelihood} can be recursively computed, for example by first iterating over generations and then over the individuals of each generation. However, the same cannot be immediately said for \eqref{output_likelihood}, where the marginalization complicates the calculation. Here we propose an iterative calculation of this likelihood that again proceeds sequentially through the tree generations and the daughter pairs of each generation. As it turns out, marginalizing over the states introduces dependencies between different daughter pairs of the same generation that result in slightly more complicated formulas. These dependencies also complicate the numerical approximation of the likelihood, but, as we will see at the end of the section, this computation can be sped up considerably by making a reasonable simplifying approximation.

Before we derive the exact formulas, we need some additional notation. Let
\[\boldsymbol{Y}^{1,\dots,i}_n:=\{\boldsymbol{Y}^1_n,\dots,\boldsymbol{Y}^i_n\}\]
denote the whole dataset of generation $n$ and
\[\mathbb{Y}_{0:n}:=\{\boldsymbol{Y}^{1,\dots,2^m}_m,m=0,\dots,n\}\]
the dataset of all generations up to generation $n$. Similarly,
\[\boldsymbol{X}^{1,\dots,i}_n:=\{\boldsymbol{X}^1_n,\dots,\boldsymbol{X}^i_n\}\]
and
\[\mathbb{X}_{0:n}:=\{\boldsymbol{X}^{1,\dots,2^m}_m,m=0,\dots,n\}.\]

To arrive at the exact formula for the likelihood, we first break up the total likelihood over the generations as follows (the dependence on $\Theta$ is suppressed to simplify the notation):
\[P(\boldsymbol{Y}_{tree})=P(\boldsymbol{Y}^1_0)\prod_{n=1}^N P(\boldsymbol{Y}^{1,\dots,2^n}_n|\mathbb{Y}_{0:n-1}).\]
Assume now that $P(\mathbb{Y}_{0:n})$ (i.e. the likelihood of the subtree consisting of the first $n$ generations) is available, and so is $P(\mathbb{X}_{0:n}|\mathbb{Y}_{0:n})$ (the state posterior over the same subtree). Consider the first two individuals of generation $n+1$, with state trajectories $\boldsymbol{X}^1_{n+1}$ and $\boldsymbol{X}^2_{n+1}$, descending from the mother cell with state trajectory $\boldsymbol{X}^1_{n}$. 

Adding the information of this daughter pair to the posterior of the previous generations, we get
\begin{align*}
&P\left(\boldsymbol{X}^1_{n+1},\boldsymbol{X}^2_{n+1},\mathbb{X}_{0:n}|\boldsymbol{Y}^1_{n+1},\boldsymbol{Y}^2_{n+1},\mathbb{Y}_{0:n}\right)=
\frac{P\left(\boldsymbol{Y}^1_{n+1},\boldsymbol{Y}^2_{n+1},\mathbb{Y}_{0:n}|\boldsymbol{X}^1_{n+1},\boldsymbol{X}^2_{n+1},\mathbb{X}_{0:n}\right)P\left(\boldsymbol{X}^1_{n+1},\boldsymbol{X}^2_{n+1},\mathbb{X}_{0:n}\right)}
{P\left(\boldsymbol{Y}^1_{n+1},\boldsymbol{Y}^2_{n+1},\mathbb{Y}_{0:n}\right)}=\\
&\frac{P\left(\boldsymbol{Y}^1_{n+1}|\boldsymbol{X}^1_{n+1}\right)P\left(\boldsymbol{Y}^2_{n+1}|\boldsymbol{X}^2_{n+1}\right)P\left(\boldsymbol{X}^1_{n+1},\boldsymbol{X}^2_{n+1}|\mathbb{X}_{0:n}\right)}
{P\left(\boldsymbol{Y}^1_{n+1},\boldsymbol{Y}^2_{n+1}|\mathbb{Y}_{0:n}\right)}
\frac{P\left(\mathbb{Y}_{0:n}|\mathbb{X}_{0:n}\right)P\left(\mathbb{X}_{0:n}\right)}{P\left(\mathbb{Y}_{0:n}\right)}=\\
&\frac{P\left(\boldsymbol{Y}^1_{n+1}|\boldsymbol{X}^1_{n+1}\right)P\left(\boldsymbol{Y}^2_{n+1}|\boldsymbol{X}^2_{n+1}\right)P\left(\boldsymbol{X}^1_{n+1},\boldsymbol{X}^2_{n+1}|\boldsymbol{X}^1_n\right)}
{P\left(\boldsymbol{Y}^1_{n+1},\boldsymbol{Y}^2_{n+1}|\mathbb{Y}_{0:n}\right)}P(\mathbb{X}_{0:n}|\mathbb{Y}_{0:n}).
\end{align*}

The normalizing constant of the above posterior extends $P(\mathbb{Y}_{0:n})$ with the daughter pair of the next generation:
\[P(\boldsymbol{Y}^1_{n+1},\boldsymbol{Y}^2_{n+1}|\mathbb{Y}_{0:n})=\iint \Big( P\left(\boldsymbol{Y}^1_{n+1},\boldsymbol{Y}^2_{n+1}|\boldsymbol{X}^1_{n+1},\boldsymbol{X}^2_{n+1}\right)P\left(\boldsymbol{X}^1_{n+1},\boldsymbol{X}^2_{n+1}|\boldsymbol{X}^1_n\right)
d\boldsymbol{X}^1_{n+1}d\boldsymbol{X}^2_{n+1}\Big)P(\boldsymbol{X}^1_n|\mathbb{Y}_{0:n})d\boldsymbol{X}^1_n\]

The above formulas allow us to update the starting posterior and likelihood with the first daughter pair from generation $n+1$. However, to add the second daughter pair (cells 3 and 4 of generation $n+1$, descending from cell 2 of generation $n$), we need to take into account the information provided by the first pair:
\begin{equation}\label{joint_posterior}
P\left(\boldsymbol{X}^{3,4}_{n+1},\boldsymbol{X}^{1,2}_{n+1},\mathbb{X}_{0:n}|\boldsymbol{Y}^{3,4}_{n+1},\boldsymbol{Y}^{1,2}_{n+1},\mathbb{Y}_{0:n}\right)=
P\left(\boldsymbol{X}^{1,2}_{n+1},\mathbb{X}_{0:n}|\boldsymbol{Y}^{1,2}_{n+1},\mathbb{Y}_{0:n}\right)
\frac{P\left(\boldsymbol{Y}^{3,4}_{n+1}|\boldsymbol{X}^{3,4}_{n+1}\right)P\left(\boldsymbol{X}^{3,4}_{n+1}|\boldsymbol{X}^{1,2}_{n+1},\mathbb{X}_{0:n}\right)}
{P(\boldsymbol{Y}^{3,4}_{n+1}|\boldsymbol{Y}^{1,2}_{n+1},\mathbb{Y}_{0:n})}.
\end{equation}
The above expression can be simplified by noting that $P\big(\boldsymbol{X}^{3,4}_{n+1}|\boldsymbol{X}^{1,2}_{n+1},\mathbb{X}_{0:n}\big)=P\big(\boldsymbol{X}^{3,4}_{n+1}|\mathbb{X}_{0:n}\big)$, i.e. daughter pairs of the same generation are conditionally independent given the parent states. However, the presence of the term $P\big(\boldsymbol{X}^{1,2}_{n+1},\mathbb{X}_{0:n}|\boldsymbol{Y}^{1,2}_{n+1},\mathbb{Y}_{0:n}\big)$ implies that, by taking into account the measurements of the first daughter pair, our posterior belief about the $n$-th generation states also needs to be updated before proceeding to the next pair. This leads to the creation of dependencies between the tree branches and means that they cannot be treated independently of each other, a feature than can create computational difficulties when one attempts to approximate the joint posterior by simulation. We thus make the \emph{simplifying assumption} that
\[P(\boldsymbol{X}^i_n|\boldsymbol{Y}^{2i-1,2i}_{n+1},\mathbb{Y}_{0:n})\approx P(\boldsymbol{X}^i_n|\mathbb{Y}_{0:n}).\]
In words, we assume that \emph{the additional state information transferred from the measurement of a daughter pair at generation $n+1$ to their corresponding mother at generation $n$ is negligible in comparison to the information provided by the previous generations to the mother}. As we will see, this allows us to treat each mother-daughter pair within a generation independently from the rest.

Continuing the analysis of the first two daughter pairs from above, we have that
\begin{align*}P\left(\boldsymbol{X}^{1,2}_{n+1},\mathbb{X}_{0:n}|\boldsymbol{Y}^{1,2}_{n+1},\mathbb{Y}_{0:n}\right)&\approx P\left(\boldsymbol{X}^{1,2}_{n+1}|\mathbb{X}_{0:n},\boldsymbol{Y}^{1,2}_{n+1},\mathbb{Y}_{0:n}\right)
P\left(\mathbb{X}_{0:n}|\mathbb{Y}_{0:n}\right)=\\
& P\left(\boldsymbol{X}^{1,2}_{n+1}|\mathbb{X}_{0:n},\boldsymbol{Y}^{1,2}_{n+1}\right)
P\left(\mathbb{X}_{0:n}|\mathbb{Y}_{0:n}\right).
\end{align*}
This fact therefore leads to a simplification of the conditional likelihood, $P(\boldsymbol{Y}^{3,4}_{n+1}|\boldsymbol{Y}^{1,2}_{n+1},\mathbb{Y}_{0:n})$:
\begin{align*}
&P(\boldsymbol{Y}^{3,4}_{n+1}|\boldsymbol{Y}^{1,2}_{n+1},\mathbb{Y}_{0:n})=\\ &\iint \Big( P\big(\boldsymbol{Y}^{3,4}_{n+1}|\boldsymbol{X}^{3,4}_{n+1}\big)P\big(\boldsymbol{X}^{3,4}_{n+1}|\boldsymbol{X}^2_n\big)d\boldsymbol{X}^{3,4}_{n+1}\Big) P\big(\boldsymbol{X}^{1,2}_{n+1}|\boldsymbol{X}^2_n,\boldsymbol{Y}^{1,2}_{n+1}\big)P\left(\boldsymbol{X}^2_n|\mathbb{Y}_{0:n}\right)d\boldsymbol{X}^{1,2}_{n+1}d\boldsymbol{X}^2_n=\\
&\iint \Big( P\big(\boldsymbol{Y}^{3,4}_{n+1}|\boldsymbol{X}^{3,4}_{n+1}\big)P\big(\boldsymbol{X}^{3,4}_{n+1}|\boldsymbol{X}^2_n\big)d\boldsymbol{X}^{3,4}_{n+1}\Big) P\left(\boldsymbol{X}^2_n|\mathbb{Y}_{0:n}\right)d\boldsymbol{X}^2_n=P(\boldsymbol{Y}^{3,4}_{n+1}|\mathbb{Y}_{0:n}),
\end{align*}
and the total likelihood of generation $n+1$ (conditioned on $\mathbb{Y}_{0:n}$) can be decomposed as a product of likelihoods over the individual daughter pairs.

Finally, the joint posterior \eqref{joint_posterior} can be also decomposed as:
\begin{equation}\label{decomp_posterior}
P\left(\boldsymbol{X}^{3,4}_{n+1},\boldsymbol{X}^{1,2}_{n+1},\mathbb{X}_{0:n}|\boldsymbol{Y}^{3,4}_{n+1},\boldsymbol{Y}^{1,2}_{n+1},\mathbb{Y}_{0:n}\right)=
P\left(\boldsymbol{X}^{1,2}_{n+1}|\boldsymbol{Y}^{1,2}_{n+1},\mathbb{X}_{0:n}\right)P\left(\boldsymbol{X}^{3,4}_{n+1}|\boldsymbol{Y}^{3,4}_{n+1},\mathbb{X}_{0:n}\right)
P\left(\mathbb{X}_{0:n}|\mathbb{Y}_{0:n}\right).
\end{equation}

These facts will be put in use in the next section, where a Sequential Monte Carlo algorithm for the approximation of the tree likelihood will be presented.

\subsection{Recursive likelihood approximation}
\label{subsec:SMC}
Our SMC scheme is used to approximate $P(\boldsymbol{Y}_{tree} \mid \Theta)$, i.e. the likelihood of a set of measurements over a tree starting from a single individual, given a set of parameters $\Theta$, under the simplifying assumption presented above. Our algorithm uses this assumption to exploit the conditional independence structure of the tree dynamics it generates in order to break down the likelihood computation. More concretely, the idea is to start at the root of the tree (i.e., a single cell) and recursively propagate the data likelihood from one generation to the next, treating each mother cell and its progeny independently. This can be understood as a generalization of recursive filtering for tree-structured data. To illustrate this better, we present the treatment of a single mother-daughter triplet in detail.

Assume that $L$ samples (particles) from the known (prior) distribution of $x_n^i(T_n^i)$ are available. First, a pair of daughter cells is generated according to the transition probabilities $P(x_{n+1}^{2i-1}(0), x_{n+1}^{2i}(0) | x_n^i(T_n^i), \Theta)$ for each particle. Given the daughters' initial conditions, we next simulate each daughter until its own division time and calculate the likelihoods $P(\boldsymbol{Y}_{n+1}^{2i-1} \mid \boldsymbol{X}_{n+1}^{2i-1,l}, \Theta)$  and $P(\boldsymbol{Y}_{n+1}^{2i} \mid \boldsymbol{X}_{n+1}^{2i,l},\Theta)$ for $l=1,\dots,L$.

By assigning to the $l$-th particle a weight
\[w_{n+1}^{i,l}=P(\boldsymbol{Y}_{n+1}^{2i-1} \mid \boldsymbol{X}_{n+1}^{2i-1,l}, \Theta)P(\boldsymbol{Y}_{n+1}^{2i} \mid \boldsymbol{X}_{n+1}^{2i,l},\Theta),\]
we next compute the marginal likelihood of the $i^{th}$ daughter pair of generation $n+1$ by averaging the weights for all the particles:
\[
P(\boldsymbol{Y}_{n+1}^{2i-1},\boldsymbol{Y}_{n+1}^{2i}|\Theta)=\frac{1}{L}\sum_{l=1}^{L} w_{n+1}^{i,l}.
\]
After normalizing the particle weights to sum up to one, we have thus obtained weighted samples from the posteriors $P(\boldsymbol{X}_{n+1}^{2i-1} \mid \boldsymbol{Y}_{n+1}^{2i-1},\Theta)$ and $P(\boldsymbol{X}_{n+1}^{2i} \mid \boldsymbol{Y}_{n+1}^{2i},\Theta)$. The samples are subsequently unweighted by resampling $L$ particles from each posterior according to the normalized weights. These samples will serve as starting points for the daughters of the next generation. The same process is repeated for the rest of the $n^{th}$ generation mothers, before moving on to generation $n+1$. This very general procedure is summarized in Algorithm \ref{SMC_tree}.\\
~\\

\begin{algorithm}[H]
 \caption{The SMC algorithm for tree likelihood calculation}
 \label{SMC_tree}
\KwResult{Estimate of $P(\boldsymbol{Y}_{tree}|\Theta)$}

Create $L$ replicates of $\boldsymbol{X}_0^1$, $\{\boldsymbol{X}_0^{1,l}\}_{l=1}^L$ \;
Set $P(\boldsymbol{Y}_{tree}|\Theta)=1$ \;

\For{n = 0 to N-1}{
    \For{i=1 to $2^n$}{
    \For{l = 1 to L}{
       Simulate a pair of daughter cells, $\boldsymbol{X}_{n+1}^{2i-1,l}$ and $\boldsymbol{X}_{n+1}^{2i,l}$ with initial conditions drawn from $ P(x_{n+1}^{2i-1}(0),  x_{n+1}^{2i}(0) | x_n^i(T_n^i), \Theta)$ \;
       Compute the weights $w_{n+1}^{i,l} = P(\boldsymbol{Y}_{n+1}^{2i-1,l} \mid \boldsymbol{X}_{n+1}^{2i-1,i }, \Theta)P(\boldsymbol{Y}_{n+1}^{2i,l} \mid \boldsymbol{X}_{n+1}^{2i,l}, \Theta)$ \;
    }
    Compute marginal likelihood of the triplet: $P(\boldsymbol{Y}_{n+1}^{2i-1},\boldsymbol{Y}_{n+1}^{2i}|\Theta) = L^{-1}\sum_{l=1}^L w_{n+1}^{i,l}$ \;
    Update tree likelihood estimate: $P(\boldsymbol{Y}_{tree}|\Theta) = P(\boldsymbol{Y}_{tree}|\Theta)P(\boldsymbol{Y}_{n+1}^{2i-1},\boldsymbol{Y}_{n+1}^{2i}|\Theta)$ \;
    Compute normalized weights: $\widetilde{w}_{n+1}^{i,l} := w_{n+1}^{i,l}/\sum_{l=1}^L w_{n+1}^{i,l}$ \;
    Resample $\{\boldsymbol{X}_{n+1}^{2i-1,l}\}_{l=1}^L$ and $\{\boldsymbol{X}_{n+1}^{2i,l}\}_{l=1}^L$ according to $\{\widetilde{w}_{n+1}^{i,l}\}_{l=1}^L$ \;
   }
}
\end{algorithm}
~\\

\paragraph{Notes:}
\begin{enumerate}
\item Since we assume that the measurements from individual trees are independent from each other, the joint likelihood of a dataset consisting of several trees is simply a product of the likelihood of the individual trees. The likelihoods of individual trees can be thus estimated in parallel. Moreover, looking at the algorithm structure for a single tree, the likelihood calculation can be parallelized at two levels: 1) the mother cells of a given generation can be treated independently of each other 2) individual particle calculations for a given mother-daughter triplet can be done in parallel.
\item When no randomness is present in the transition mechanism from the mother to the daughters, the daughter cells can be treated completely independently given the state of the mother. This implies that they can be assigned independent weights at line 7 of Algorithm \ref{SMC_tree} : $w_{n+1}^{2i-1,l} = P(\boldsymbol{Y}_{n+1}^{2i-1,l} | \boldsymbol{X}_{n+1}^{2i-1,i }, \Theta)$ and $w_{n+1}^{2i,l} = P(\boldsymbol{Y}_{n+1}^{2i,l} | \boldsymbol{X}_{n+1}^{2i,i }, \Theta)$. In this way, the marginal likelihood computation of line 9 can be written as $P(\boldsymbol{Y}_{n+1}^{2i-1},\boldsymbol{Y}_{n+1}^{2i}|\Theta) = \left(L^{-1}\sum_{l=1}^L w_{n+1}^{2i-1,l}\right)\left( L^{-1}\sum_{l=1}^L w_{n+1}^{2i,l}\right)$. Additionally, the normalized weights for the particle populations of the two daughters (line 11) are now independent: $\widetilde{w}_{n+1}^{2i-1,l} := w_{n+1}^{2i-1,l}/\sum_{l=1}^L w_{n+1}^{2i-1,l}$ and $\widetilde{w}_{n+1}^{2i,l} := w_{n+1}^{2i,l}/\sum_{l=1}^L w_{n+1}^{2i,l}$, which finally implies that the populations $\{\boldsymbol{X}_{n+1}^{2i-1,l}\}_{l=1}^L$ and $\{\boldsymbol{X}_{n+1}^{2i,l}\}_{l=1}^L$ (line 12) can be resampled independently.
\item The most computationally intensive step of the algorithm lies between lines 5-9, where the marginal likelihood of each mother-daughter triplet needs to be computed. Depending on the type of the unobserved state dynamics, accurate marginalization may require the use of very large particle numbers and greatly increase the computational cost of the algorithm. Typically, the situation is worse when the hidden state contains components driven by stochastic dynamics. This challenge has already been recognized and addressed in the literature, since it also appears in the parameter inference problem from independent single-cell trajectories \cite{wilkinson2011,Amrein11}. One can thus employ one of the several available alternatives at this step, such as sequential computation of the likelihood \cite{wilkinson2011}, or the use of approximating dynamics \cite{wilkinson2011,Stathopoulos11,Wilkinson05}.

\end{enumerate}

\subsection{A pseudo-marginal MCMC sampler for parameter inference}
\label{subsec:inference}
\begin{figure}[H]
    \centering
    \includegraphics[width=5.0in]{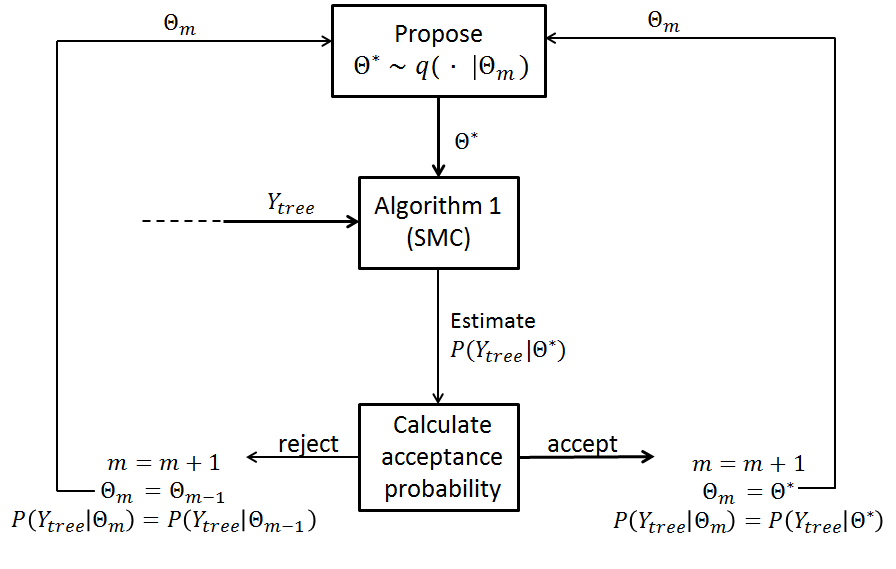}
    \caption{An overview of our inference framework}
    \label{fig:framework}
\end{figure}

Let us denote with $\overline{\Theta}$ the set of all unknown parameters from the vector $\Theta$ that we would like to infer. The goal of Bayesian inference is to compute or approximate via sampling the posterior distribution of these parameters, $P(\overline{\Theta}|\boldsymbol{Y}_{tree})\propto P(\boldsymbol{Y}_{tree}|\overline{\Theta})\pi(\overline{\Theta})$. To this end, we follow the ``pseudo-marginal'' MCMC approach \cite{Andrieu09}, according to  which  a Markov Chain Monte Carlo (MCMC) sampler makes use of the noisy marginal likelihood estimates provided by the SMC algorithm of Section \ref{subsec:SMC} to generate samples from the posterior of $\overline{\Theta}$. Crucially, despite the presence of noise in the SMC-based likelihood estimation, the MCMC sampler is still able to target the correct distribution. This property is in fact guaranteed for any marginal likelihood estimator that provides unbiased estimates of the likelihood \cite{Andrieu09}, although the mixing properties of the MCMC chain will of course depend on the estimate variability. Given the fact that the SMC algorithm employed in this work indeed provides unbiased marginal likelihood estimates \cite{Kuensch13}, the pseudo-marginal MCMC scheme described in Algorithm \ref{algorithm:framework} targets the correct posterior parameter distribution. In addition to the pseudo code presented, we summarize our inference framework in Figure \ref{fig:framework}.
~\\

\begin{algorithm}[H]
 \caption{The pseudo-marginal MCMC sampler for parameter inference}
 \label{algorithm:framework}
\SetAlgoLined
\KwResult{\{$\overline{\Theta}_m\}_{m=1}^M\sim P(\overline{\Theta}| \boldsymbol{Y}_{tree})$}

Draw an initial point $\overline{\Theta}_1$ from the prior $\pi(\overline{\Theta})$\;
Estimate the likelihood $P(\boldsymbol{Y}_{tree} |\overline{\Theta}_1)$ using Algorithm \ref{SMC_tree}\;

\For{$m = 2$ to $M$}{
     Propose a parameter vector $\overline{\Theta}^*$ according to a proposal distribution $q(\overline{\Theta}^* |\overline{\Theta}_{m-1})$\;
     Calculate the likelihood $P(\boldsymbol{Y}_{tree} \mid \overline{\Theta}^*)$ using Algorithm \ref{SMC_tree} \;
     Sample $u\sim\mathcal{U}([0,1])$\;
     If \[u<min\{1, \frac{P(\boldsymbol{Y}_{tree} \mid \overline{\Theta}^*) \cdot q(\overline{\Theta}_{m-1}|\overline{\Theta}^*)}{P(\boldsymbol{Y}_{tree} \mid \overline{\Theta}_{m-1}) \cdot q(\overline{\Theta}^*|\overline{\Theta}_{m-1})}\},\]
     accept the proposed parameters and set $\overline{\Theta}_m = \overline{\Theta}^*$ and $P(\boldsymbol{Y}_{tree} \mid \overline{\Theta}_{m})=P(\boldsymbol{Y}_{tree} \mid \overline{\Theta}^*)$; else, set $\overline{\Theta}_m = \overline{\Theta}_{m-1}$ and $P(\boldsymbol{Y}_{tree} \mid \overline{\Theta}_{m})=P(\boldsymbol{Y}_{tree} \mid \overline{\Theta}_{m-1})$.
}
\end{algorithm}
~\\

It is important to note that, in order for the above sampler to target the correct posterior despite the noisy likelihood estimates, it is necessary to keep the \emph{previous} estimate of the likelihood in the acceptance ratio whenever a proposed parameter vector is rejected, and not get another SMC estimate for the same point \cite{Andrieu09}. Fortunately, this also reduces the computational burden of the sampler, as it requires one likelihood estimation per iteration. On the negative side, the use of very noisy SMC estimates may considerably slow down the mixing of the sampler, since the chain may get trapped at a point with artificially large likelihood value. However, as the variance of the estimator decreases (e.g. through the use of larger particle sample sizes), it is expected that the mixing speed of our sampler will converge to that of a sampler with perfect (i.e. noiseless) marginal likelihood information.

\section{Applications and results}
\label{sec:applications}

In the following sections we will consider two possible examples for the dynamical system $S$ and demonstrate the application of our inference method on these cases. In both examples, we assume that a cell is characterized by a discrete state, $x_d$. Over time and across generations, cells stochastically adopt a certain \emph{type} determined by $x_d$. The cell type in turn determines the evolution of a continuous state vector, $x_c$, which may for example correspond to the immature and mature molecule types of a fluorescent reporter. In abstract terms, $x_d$ may be thought of as the state a gene, whose activity affects the cell phenotype.

In the first example, the discrete state dynamics is described  by a generalized two-type branching process. According to this scenario, the type of a cell is fixed throughout its lifetime and may change only at cell division, since the types of the daughters depend probabilistically on the type of the mother. In the second example, the cell type may stochastically vary throughout the cell lifetime according to a two-state continuous-time Markov chain, while the two daughters are assumed to inherit the type of the mother. To test the performance of our inference framework, we generated simulated datasets for the two example systems and used them to infer parameters of interest in each case. Details about some of the parameters used in the data generation process are provided in Table \ref{table:data_generation}. The results for each example system are summarized below.

\begin{table}[H]
\caption{Parameters used in the synthetic data generation}
\begin{tabular}{c|c|c}
{\bf General}        & {\bf Example 1}    & {\bf Example 2}    \\  
Number of trees               &2                        & 1                   \\
Number of generations               & 5                       & 5                   \\
Cell division time ($min$)              & 30                       & 30                  \\
Measurement interval ($min$)              & 5                        & 5                  \\
{\bf Measurement model related}               &                          &                         \\ 
GFP production rates for OFF type $\alpha_{OFF}  (a.u. min^{-1})$               &           0.2              &            0.05             \\
GFP production rates for ON type $\alpha_{ON} (a.u. min^{-1})$               &          1              &           20             \\
GFP maturation rate $m  (min^{-1})$               &           0.0462               &            0.0462             \\
GFP dilution rate $d (min^{-1})$               &                   0.0261       &         0.0231                \\
GFP to fluorescence scaling constant $c$               &                  100      &         100               \\
measurement variance $\sigma^2$               &                  500      &         500               \\
\end{tabular}
\label{table:data_generation}
\end{table}

\subsection{ Example 1: A two-type branching process with dynamic readouts}
\label{subsec:example1}
\begin{figure}[H]
    \centering
    \includegraphics[width=5.0in]{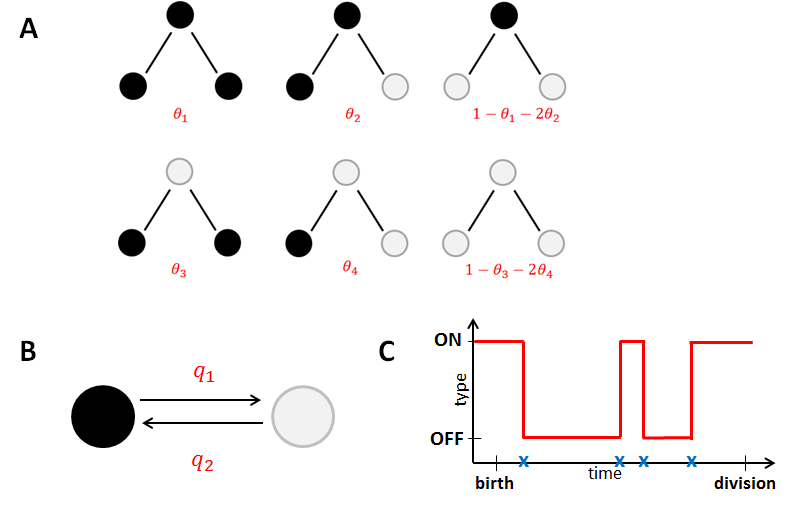}
    \caption{A.) \textbf{Example 1}All possible daughter pairs from a single mother and the corresponding probabilities of obtaining those pairs to be inferred. The different cell types here are black and white (OFF and ON). Note that the probability of obtaining first daughter cell black and the second white is same as obtaining the first daughter cell white and the second black. B.) \textbf{Example 2:} Cell types switch during the cell lifetime according to a two state CTMC with rates $q_1$ and $q_2$, C.)  An example of the evolution of a cell type during its lifetime. The blue crosses on the time axis indicate switching point from ON-OFF or vice versa.}
    \label{fig:parameters}
\end{figure}

In this example, cells can adopt one of two possible types (ON or OFF) and maintain their type throughout their lifetime, which, for simplicity, we assume to be the same and equal to $T$ for every cell. At cell division, the daughter cell types are determined based on the type of the mother cell, according to a set of transition probabilities, as illustrated in Figure \ref{fig:parameters}A. In turn, the type of each cell is assumed to determine the production rate of a fluorescent reporter protein (such as GFP) which can then be observed using fluorescence time-lapse microscopy.

The state vector of each cell is thus defined as $x=\begin{bmatrix}x_d &G_{imm}&G_{mat}\end{bmatrix}$, where $x_d$ contains the cell type, while $G_{imm}$ and $G_{mat}$ correspond to the concentrations of the immature (dark) and mature (fluorescent) forms of the fluorescent reporter. Out of these, we assume that we can only obtain noisy measurements of $G_{mat}$ at discrete points in time. Contrary to the cell type, the concentrations of the two reporter species are carried over from the mother to the daughters unchanged. That is, $G_{imm,n+1}^{2i-1}(0) = G_{imm,n}^{i}(T)$, $G_{mat,n+1}^{2i-1}(0) = G_{mat,n}^{i}(T)$ and similarly for the second daughter. This is a reasonable modeling assumption, given that a daughter cell has half the volume of the mother and receives roughly half of its protein content as well.

The fluorescent reporter dynamics of every cell evolves according to the following set of linear ODEs:

\begin{align}
\dot{G}_{imm} &= \alpha(x_d) - \delta \cdot G_{imm} - m \cdot G_{imm}\\
\dot{G}_{mat} &= - \delta \cdot G_{mat} + m \cdot G_{imm},
\end{align}
where $\alpha(x_d),~\delta$ and $m$ are reporter protein production, dilution and maturation rates respectively. The production rate is determined by the cell type: for an OFF-type cell, $\alpha(OFF)=\alpha_{OFF}$, while a cell of the ON type has $\alpha(ON)=\alpha_{ON}>\alpha_{OFF}$.

As described above, we assume that noise-corrupted measurements proportional to the $G_{mat}$ species are available at $M$ points, $t_1, ..., t_M$, during the life of every cell. The readout of a single cell at a given time is therefore assumed to be a scaled and noisy version of the $G_{mat}$ concentration:

\begin{equation}
y(t_m) \propto \mathcal{N}(c \cdot G_{mat},\sigma^2 \cdot G_{mat}),
\label{eq:fluorescence}
\end{equation}
where $c$ and the measurement variance $\sigma^2$ are known. Given that individual measurements for each cell are independent from each other, the expression for the likelihood $P(\boldsymbol{Y}_n^{i} \mid \boldsymbol{X}_{n}^{i}, \Theta)$, where $\boldsymbol{Y}_n^i=\{y_n^i(t_m),~m=1,\dots,M\}$, is given by

\begin{equation}
P(\boldsymbol{Y}_n^i \mid \boldsymbol{X}_n^i) = \prod_{m = 1}^{M}  P(y_n^i(t_m) \mid G_{mat,n}^i(t_m)).
\end{equation}

Using this type of reporter measurements for every cell belonging to a fully observed tree spanning $N$ generations, our goal is to infer: 1. The transition probabilities that govern cell type switching ($\theta_1,\dots,\theta_4$ in Fig. \ref{fig:parameters}A) and 2. The type-specific reporter production rates $\alpha_{OFF}$ and $\alpha_{ON}$.

If the type of each cell was readily measurable, the use of simple maximum likelihood estimators for branching processes would suffice to obtain all the necessary discrete state statistics from a fully observed tree, making the use of the reporter model unnecessary. However, the intervening reporter maturation step, the slow dilution dynamics and the sparse, noisy sampling, make inference much more challenging and require the use of the sophisticated computational machinery presented in this work.

To test the performance of our algorithm on this system we simulated dataset in which we generated the daughter types according to the transition probabilities shown in Table \ref{example1_transition}, with GFP reporter production rates $\alpha_{OFF}=0.2 \quad a.u. min^{-1}$ and $\alpha_{ON}=1 \quad a.u. min^{-1}$. 

\begin{table}[h!tb]
\centering
\begin{tabular}{c c|c|c|c|c|}
\cline{3-6}
& &\multicolumn{4}{c|}{Daughter Types}\\ \cline{3-6}
& &(OFF,OFF) & (OFF,ON) & (ON,OFF) & (ON,ON) \\
\cline{1-6}
\multicolumn{1}{|c|}{\multirow{2}{*}{Mother type}} & OFF & $\theta_1=0.6$ & $\theta_2=0.1$ & $\theta_2$ & $1-\theta_1-2\theta_2$ \\
\cline{2-6}
\multicolumn{1}{|c|}{} & ON & $\theta_3 =0.1$ & $\theta_4=0.05$ & $\theta_4$ & $1-\theta_3-2\theta_4$\\
\cline{1-6}
\end{tabular}
\caption{Transition probabilities for the cell types considered in Example 1 and depicted on Figure \ref{fig:parameters}.}
\label{example1_transition}
\end{table}
Note that due to symmetry, the second and third entries of each row are equal. Moreover, the values of the first and second entries in each row determine the rest of the entries, since every row sums to one. We therefore considered $\theta_1$, $\theta_2$, $\theta_3$ and $\theta_4$ as unknown, together with the reporter production rates at each state, $\theta_5:=\alpha_{OFF}$ and $\theta_6:=\alpha_{ON}$. For all unknown parameters, we considered flat priors supported on the appropriate sets ($[0,1]$ in the case of transition probabilities, and $[0,+\infty]$ for the production rates). The rest of the system parameters were fixed at the values reported in Table \ref{table:data_generation}.

We ran the pseudo-marginal MCMC sample (Algorithm \ref{algorithm:framework}) to generate samples from the posterior distribution of $\overline{\Theta} =[\theta_1~\theta_2~\theta_3~\theta_4~\alpha_{OFF}~\alpha_{ON}]$, using appropriate proposal densities for the different types of parameters: transition probabilities $\theta_1$ and $\theta_2$ were sampled from the three-dimensional simplex with the help of a Dirichlet distribution with its mode located at the current parameter values. Similarly, $\theta_3$ and $\theta_4$ were sampled independently from a second Dirichlet distribution, whereas $\theta_5$ and $\theta_6$ were generated from independent log-normal distributions with log-means equal to $\log(\theta_5)$ and $\log(\theta_6)$ respectively, and log-standard deviation equal to 0.02. The estimated posterior distributions $P(\overline{\Theta}|Y)$ based on 1500 MCMC steps are given in Figure \ref{fig:Posteriors}, where it can be clearly seen that the inferred posterior means (black dashed lines) are located close to the true parameter values (red lines).

\begin{figure}[H]
    \centering
    \includegraphics[width=5.0in]{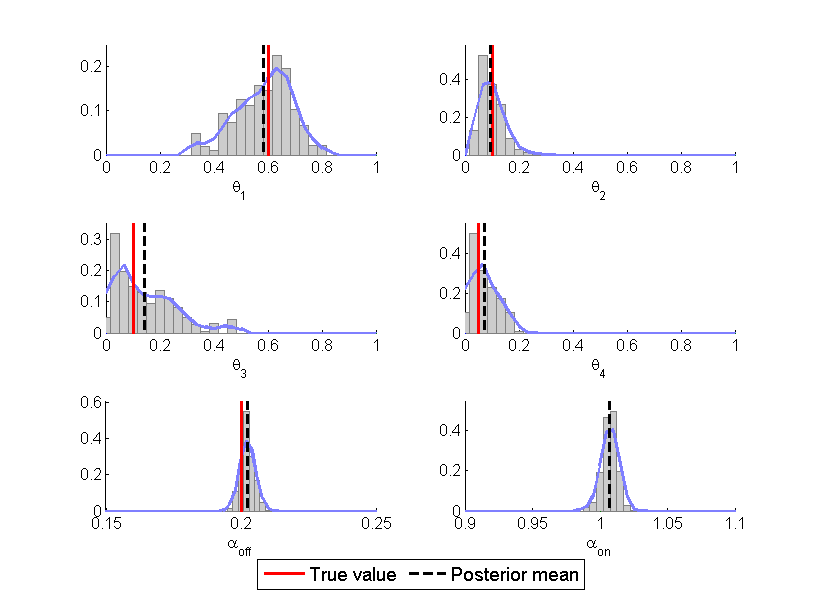}
    \caption{Histograms of sampled posterior distributions for the six unknown parameters (after the burn-in of the first 300 iterations had been discarded). The red vertical bar is positioned at the true parameter value (the one used for data generation), while the black dashed line is positioned at the estimated posterior mean (they are overlapping in the last plot). The blue curves are obtained by smoothing of the histograms.}
    \label{fig:Posteriors}
\end{figure}

It is known from theory that the SMC estimator for the likelihood approximation is unbiased \cite{Kuensch13}. To assess its variability as a function of the number of particles and thus determine approximately the particle number required for sufficiently accurate likelihood estimation in the MCMC sampler, we estimated the likelihood of a small tree $P(Y_{tree}|\overline\Theta)$ with different numbers of particles given the true parameter values $\overline\Theta_{true}$. As can be seen on Figure \ref{fig:tiger}, the average of 100 likelihood estimates fluctuates considerably for small particle number, but stabilizes as the particle number increases. With the increase of the number of particles the coefficient of variation of the likelihood calculations also drops quickly.

\begin{figure}[H]
    \centering
    \begin{subfigure}[b]{0.47\textwidth}
        \includegraphics[width=\textwidth]{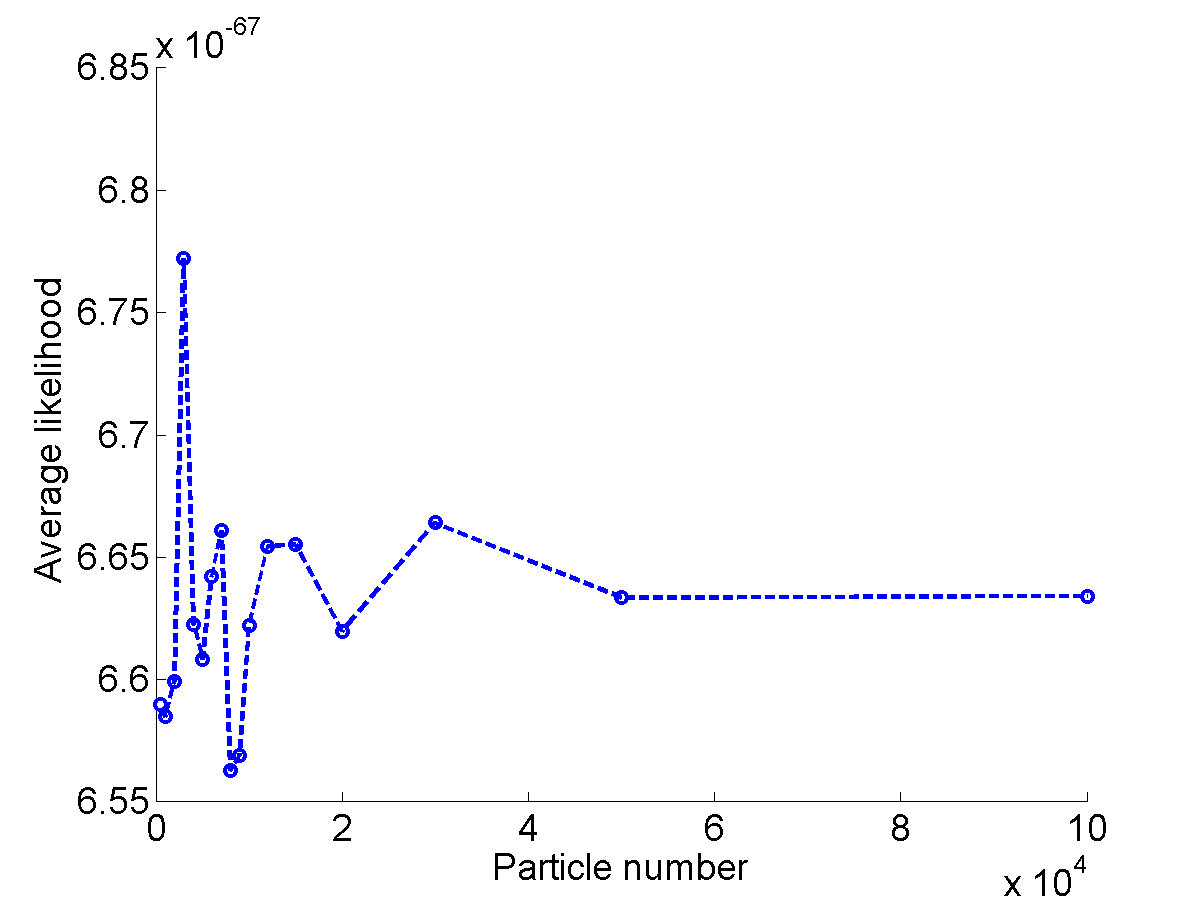}
        \label{fig:gull}
    \end{subfigure}
    \begin{subfigure}[b]{0.47\textwidth}
        \includegraphics[width=\textwidth]{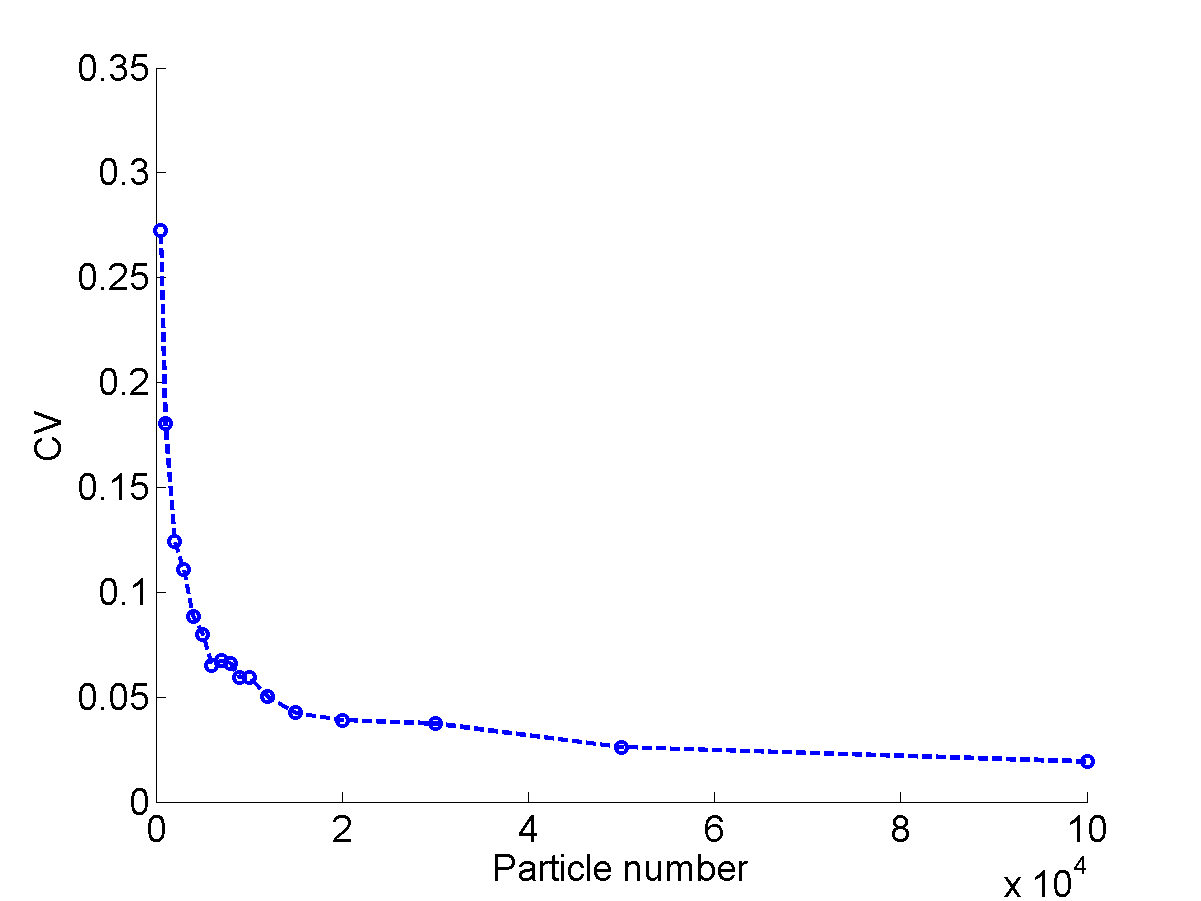}
        \label{fig:tiger}
    \end{subfigure}
    \caption{Mean of 100 estimates of $P(Y_{tree} | \overline\Theta_{true})$ with SMC vs. number of particles used (left) and their CV (right).}\label{fig:tiger}
\end{figure}

\subsection{ Example 2: Stochastic cell type switching}
In the second example, we assume that the cell type evolves according to a two-state continuous-time Markov chain (CTMC) with rates $q_1$ and $q_2$ for the OFF-to-ON and ON-to-OFF transitions respectively. At division, each daughter inherits the type of its mother (together with the reporter concentrations, as before), but subsequently evolves independently from other cells according to the CTMC dynamics, as shown on Figure \ref{fig:parameters}B and C. In this case, the reporter production rate alternates between $\alpha(ON)$ and $\alpha(OFF)$ in accordance with the cell type.

Using the same type of reporter model dynamics and readouts as in the previous example, our goal in this case it to infer the CTMC transition rates, $q_1$ and $q_2$ (Figure \ref{fig:parameters} B ), while assuming the rest of the parameters known.

While in the first example system the tree-structure information was essential for inference as the cell types can only change at cell division, it is not equally obvious that our method will outperform traditional inference on independent single cell trajectories in this example system, where each daughter inherits the state of its mother and then evolves independently. To verify this, we additionally performed parameter inference by breaking up the tree into individual cell trajectories and considering each cell independently from the others, as it is usually done in conventional inference methods for single-cell data.

For this second type of inference, the priors for the reporter initial conditions of the daughters were determined according to the first measurement point of each cell, by discarding the part of its trajectory between the mother division and the first measurement. More specifically, the initial mature reporter ($G_{mat}$) value for each particle was extracted from the first available measurement point, by dividing the fluorescence with the scaling constant $c$. The immature reporter value ($G_{imm}$) was then drawn independently for each particle from a log-normal distribution with log-mean equal to $\log(G_{mat} \cdot\frac{d}{m})$ and appropriately tuned log-variance. With respect to the discrete states, a uniform prior on the two cell types was assumed.

For the MCMC sampler, we used two independent log-normal proposal kernels with log-means equal to the logarithm of the current parameter values at each step, with standard deviation of 0.2. Figure \ref{fig:Posteriors} compares the posterior distributions of the parameters for different datasets, obtained when the inference was done both on trees and on individual trajectories. 

\begin{figure}[H]
    \centering
    \includegraphics[width=5.0in]{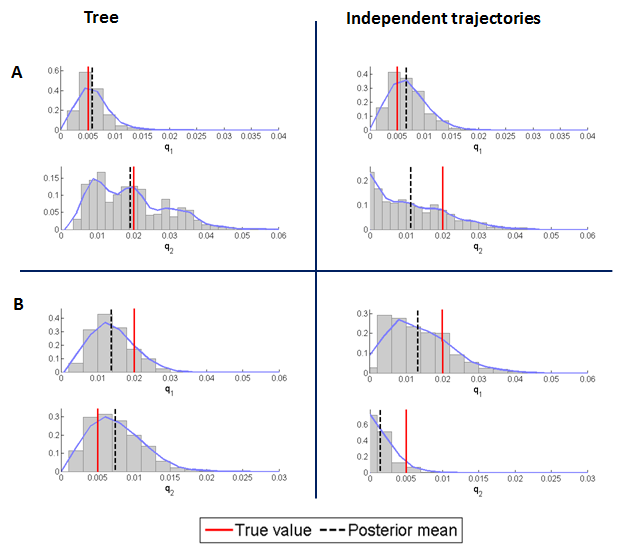}
    \caption{Histograms of sampled posterior distributions of the two unknown switching rates in Example 2 ($q1$: switching rate from OFF to ON and $q2$: switching rate from ON to OFF). The inference results based on tree-structured data (left column) are compared with the results obtained from independent cell trajectories (right column), as described in the text. Simulated data was generated using different parameter sets in panels A. and B. The red bars are positioned at the true parameter values (i.e. the ones used for data generation), while the black dashed lines indicate the estimated posterior mean. The blue curves are obtained by smoothing of the histograms.}
    \label{fig:Posteriors}
\end{figure}

One of the advantage of using tree-structured data is that the uncertainty regarding the initial conditions of each cell is greatly reduced, since the prior of each daughter state is based on the posterior of its mother. On the contrary, when the cell trajectories are assumed to be independent, the state of every cell has to be independently initialized according to the assumed prior. Moreover, when the switching rates are very small (corresponding to mean holding times that are close to or exceed the lifetime of a single cell), inference based on single-cell trajectories tends to underestimate them. This happens because information on the Markov chain state is only available during the lifetime of a cell. This is clearly not sufficient for precise estimation of the switching rates, which can only be upper-bounded by the data in this case. 

In the case of tree-structured data on the other hand, the daughter cells inherit the state of their mothers, which implies that the history of a cell's lineage is taken into an account in the inference. Practically, this is equivalent to observing coupled sample paths of the Markov chain over longer time scales, which can clearly provide more information about the switching rates. On the contrary, when the switching rates are larger and correspond to mean holding times comparable to or shorter than the lifetime of a cell, inference on independent trajectories performs similarly to the inference on trees.

\section{Discussion}
\label{sec:conclusions}

In this work we proposed a parameter inference method for stochastic single-cell dynamics from tree-structured data. More specifically, we considered a class of systems with one or more unobserved states and fluorescent reporter readouts, observed through time-lapse microscopy, which allows tracking individual cells and their progeny over time. Our goal was to estimate the posterior distribution of the unknown system parameters given such readouts. To calculate the likelihood of the data for a given parameter set, the hidden state trajectories had to be integrated out. This marginalization was accomplished with the help of a Sequential Monte Carlo (SMC) method, which recursively computes a sampling-based estimate of this analytically intractable quantity. To sample the system parameter space, we employed an MCMC scheme, which is guaranteed to target the correct parameter posterior despite the noisy likelihood estimates. The application of our method to two simple examples showed that it can correctly infer the parameters of interest and approximate their posterior distribution. Our inference framework extends to more complex applications in a straightforward manner, although its computational complexity increases with the state and parameter dimensionality.
In particular, more complex systems require a larger number of particles to achieve a reasonable accuracy of the SMC-based likelihood estimator. 
If the latter is too noisy, one may observe slow mixing of the MCMC sampler and in turn poor posterior estimates. Currently, our inference algorithm is implemented in Matlab and long runs with many particles are computationally very expensive. Encoding of the algorithm into more powerful programming languages and parallelization of the sampling process can dramatically increase its performance and enable its application to more complex models and larger datasets.

Our inference framework could be very useful in the case of systems where accurate tracking of single-cell dynamics across cell lineages plays and important role. These are, for example, systems involved in stem cell fate decisions, or stochastic phenotype switching in bacteria \cite{marjan},\cite{pap-pili}. In many such cases, stochastic fluctuations of key factors over long timescales and/or stochastic events taking place at cell division create strong mother-daughter and daughter-daughter correlations that play a crucial role in determining the overall behavior of a colony. In such cases, treatment of the measured single-cell trajectories independently from each other will result a large loss of information and biased parameter estimates. We believe that proper incorporation of the population lineage information into the parameter inference problem will thus provide the right framework for treating this type of systems and may reveal important insights into their function.


\bibliographystyle{plain}
\bibliography{references}

\clearpage{\pagestyle{empty}\cleardoublepage} 

\end{document}